# Optical vortices enabled by structural vortices

Yuanfeng Liu[1, †], Le Zhou[1, †], Mengfan Guo[1], Zongqi Xu[1], Jing Ma[1], Yongzheng Wen[1], Natalia M. Litchinitser[2], Yang Shen[1, *], Jingbo Sun[1, *] and Ji Zhou[1, *]

[1] School of Materials Science and Engineering, Tsinghua University, Beijing, 100084, China

[2] Department of Electrical and Computer Engineering, Duke University, Durham, North Carolina 27708, USA

† These authors contributed equally to this work.

* To whom the correspondence should be addressed:

shyang_mse@mail.tsinghua.edu.cn, jingbosun@tsinghua.edu.cn,

zhouji@tsinghua.edu.cn.

## Abstract

The structural symmetry of solids plays an important role in defining their linear and nonlinear optical properties. The quest for versatile, cost-effective, large-scale, and defect-free approaches and materials platforms for tailoring structural and optical properties on demand has been underway for decades. We experimentally demonstrate a bottom-up self-assembly-based organic engineered material comprised of synthesized molecules with large dipole moments that are crystallized into a spherulite structure. The molecules align in an azimuthal direction, resulting in a vortex polarity with spontaneously broken symmetry leading to strong optical anisotropy and nonlinear optical responses. These unique polarization properties of the judiciously designed organic spherulite combined with the symmetry of structured optical beams enable a plethora of new linear and nonlinear light-matter interactions, including the generation of optical vortex beams with complex spin states and on-demand topological charges at the fundamental, doubled, and tripled frequencies. The results of this work are likely to enable numerous applications in areas such as high-dimensional quantum information processing, with large capacity and high security. The demonstrated spherulite crystals facilitate stand-alone micro-scale devices that rely on the unique micro-scale spontaneous vortex polarity that is likely to enable future applications for high-dimensional quantum information processing, spatiotemporal optical vortices, and a novel platform for optical manipulation and trapping.

Dielectric polarization properties of solids have been studied for a century aiming at elucidating the underlying mechanisms of dielectric responses in organic and inorganic solids[1-4]. The polarization mechanisms in inorganic solids such as ionic crystals, rely on their crystalline structures, which determine the sizes and the arrangements of the dipoles and often limit the material's dielectric and optical properties[5]. Indeed, many fundamental physical properties such as piezoelectricity or second harmonic generation (SHG), originating from spontaneous polarization, have only been found in a few materials that possess a non-central symmetry. On the other hand, in organic materials, the influence of the crystalline lattice is significantly reduced. Instead, the dipoles are organized at the much larger, molecular level resulting in much larger dipole moments, and thus stronger interactions, and potentially providing more degrees of freedom for the materials design and intriguing ways to achieve properties that are difficult to attain or even fundamentally unavailable in inorganic crystals.

Liquid crystals with ferroelectric nematic phases [6-18] that possess spontaneous polarization have attracted significant attention owing to their unique physical properties and a wide range of photonic and electronic applications, including nonlinear frequency conversion[9, 10], non-volatile memory devices, and binary information storage[16-18]. The idea that a nematic phase could be ferroelectric was first discussed by Born in 1916. The phenomenon of polar ordering in nematic liquid crystals was first put forward by Debye and Born, who predicted that if liquid crystals were properly designed, their molecules could spontaneously organize into a polar-ordered state. In this work, we experimentally demonstrate a bottom-up assembly-based engineered material comprised of synthesized molecules with large dipole moments that are crystallized into a spherulite structure. The molecules align into an azimuthal direction, resulting in a vortex polarity with spontaneously broken symmetry leading to strong optical anisotropy and nonlinearity, which greatly alters the light-matter interactions, such as the spin-orbital coupling, second harmonic generation (SHG), and sum frequency generation (SFG). Finally, we show that the vortex polarity can be used to generate a family of optical vortex beams with various polarization states and topological charges at three different wavelengths by using a single beam interaction with the spherulite crystal, offering remarkable opportunities for the next generation of information processing technologies, classical and quantum communications, and beyond.

The target molecule named 4-((4-nitrophenoxy)carbonyl)-3-(trifluoromethyl)phenyl 2-

fluoro-4-methoxybenzoate was synthesized with three carefully chosen precursors: 2-fluoro-4-methoxybenzoic acid ($C_8H_7FO_3$), 4-hydroxy-2-(trifluoromethyl)benzoic acid ($C_8H_5F_3O_3$), and 4-nitrophenol ($C_6H_5NO_3$). Figure 1a shows the spherulite molecule. The three benzene rings have been joined together with ester groups (-COO-). The pendant groups (–F, -CF$_3$, and -NO$_2$) are used to increase the polarity of the spherulite molecule. Due to the large dipole moment, the spherulite molecule may show the polar nematic order with spontaneous polarization, which is the ferroelectric nematic phase. Moreover, we choose Nitro-group as the polar group on the right end, which can further augment the polarity of the entire molecule. Figure 1b shows the electrostatic potential electron density isosurface, which suggests that the molecule can be considered as an electric dipole and the dipole moment was calculated to be 12 Debye.

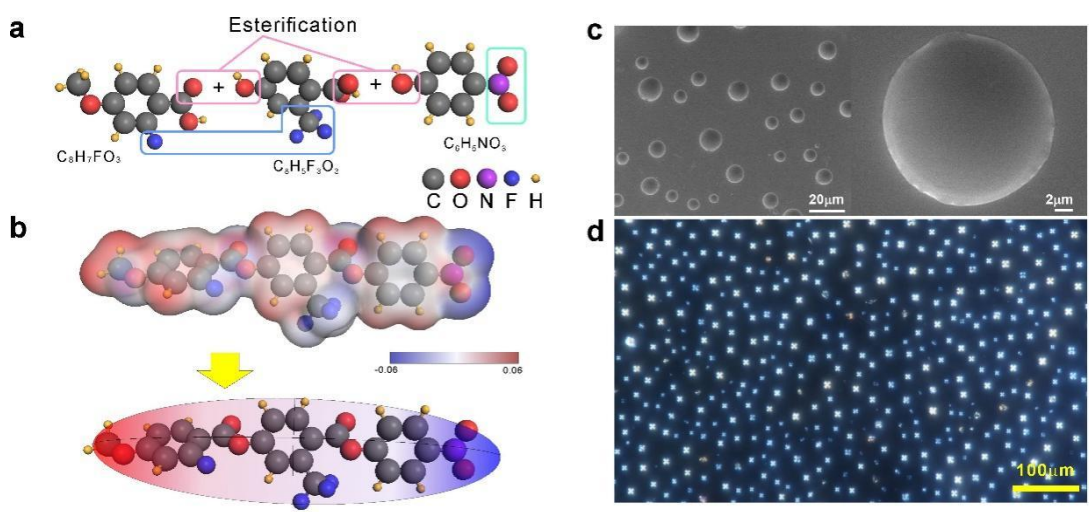

**Fig. 1 Spherulites formed by polar molecules: a**, The molecule dipole 4-((4-nitrophenoxy)carbonyl)-3-(trifluoromethyl)phenyl 2-fluoro-4-methoxybenzoate is synthesized by the precursors $C_8H_7FO_3$, $C_8H_5F_3O_3$, and $C_6H_5NO_3$ through esterification, as depicted by the pink boxes. The pendant groups in the blue (–F and -

$CF_3$) and green boxes (-$NO_2$) are used to increase the polarity of the overall molecule. **b**, Electrostatic potential electron density isosurface of the molecule, and thus it can be treated as a dipole. **c**, The SEM image of the spherulites. **d**, The optical observation of the spherulites under the polarized optical microscope.

The vortex polarity was achieved through a spherulite crystallization process, with all the polar molecules arranged azimuthally. Under 130°C, the synthesized material was fully melted, forming spherical cap-shaped droplets on a glass substrate. Each droplet was then crystallized into one single spherulite also in the spherical cap shape through a natural cooling process to room temperature. The scanning electron microscope (SEM) images in Fig. 1c show the formed spherulites with very smooth surfaces. White light interferometry and the Atomic Force Microscope (AFM) measurements revealed that all the produced spherulites have almost the same curvature despite different sizes, as shown in Supplementary Fig. S2 and S3. Differential Scanning Calorimetry (DSC) measurement was used to determine the phase transition point from the ferroelectric nematic phase to a conventional nematic phase, which is around 82°C. Detailed DSC can be found in Supplementary Fig. S4. Figure 1d shows an image obtained using a polarized optical microscope, revealing the characteristic "Maltese cross" pattern in each spherulite, which is the signature of the crystalline structure with circular symmetry[19-21]. Therefore, in what follows we assign the components of the anisotropic refractive indices, $n_r$ and $n_\theta$ corresponding to the radial and azimuthal directions, respectively.

To further determine the polarity structure, angle-resolved in-plane piezo-response force microscopy (IP-PFM) has been performed on one spherulite to investigate the underlying topology of local polarization (Fig. 2a and b). By rotating the sample in the clockwise direction by 0°, 45°, 90°, and 135°, it is found that the pattern does not change as a function of angle, confirming the rotationally symmetric polar topology in the spherical cap-shaped spherulite. In the domain pattern for each rotation, two out-of-phase semi-circle-shaped domains are separated by a vertical domain wall. We note that the detectable polar axes for each IP-PFM image in Figs. 2a and b were aligned along the vertical directions, and therefore the domain walls of low IP-PFM amplitude should be where local polarization is horizontally oriented. Considering the rotation symmetry of the structure, we conclude that the spherulite shows a signature of an azimuthally polar nematic structure with all the molecules aligned along the azimuthal direction and parallel to the substrate—the clear signature of the vortex polarity. While such vortex domains have previously been observed in the nanoscale ferroelectric arrays [23], to the best of our knowledge, here we report the first observation of this phenomenon on the microscale.

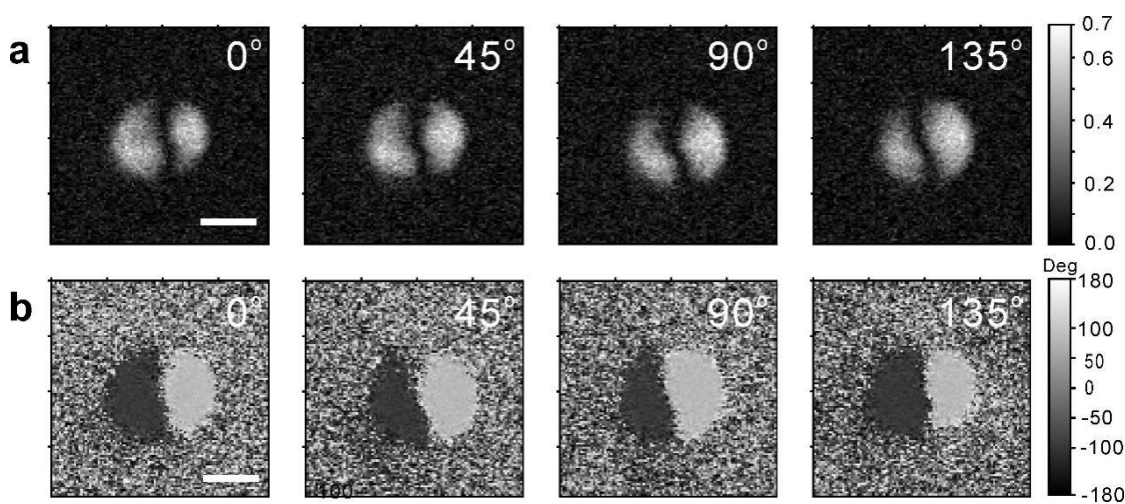

**Fig. 2 In-plane piezo-response force microscopy characterization on the**

**spherulite: a**, The PFM amplitude. **b**, The PFM phase. The scale bar in each panel is 500nm.

It is noteworthy that the vortex polarity corresponds to a non-central symmetric crystalline structure that gives rise to second-order nonlinearity. Indeed, using a 35 fs ultra-fast laser system at a wavelength of 1550 nm, we can observe the second harmonic generation (SHG) at 775nm and the third harmonic generation (THG) at 516.6nm in single a spherulite, whose morphology was characterized by AFM in Fig. 3a. The two-lobe-shaped output intensity pattern of the SHG with a dark line in the middle (along the incident linear polarization), shown in Fig. 3b, indicates that the SHG is also anisotropic, due to the azimuthal structure of the spherulite. The second-order nonlinear coefficient $d_{eff}$ can be determined by fitting the intensities of the fundamental wave and that of the second harmonic component. According to the fitting, the second-order nonlinearity is estimated to be 8.5 pm/V, approximately 1/4 of the $LiNbO_3$. Detailed fitting and calculations can be found in Supplementary Fig. S5.

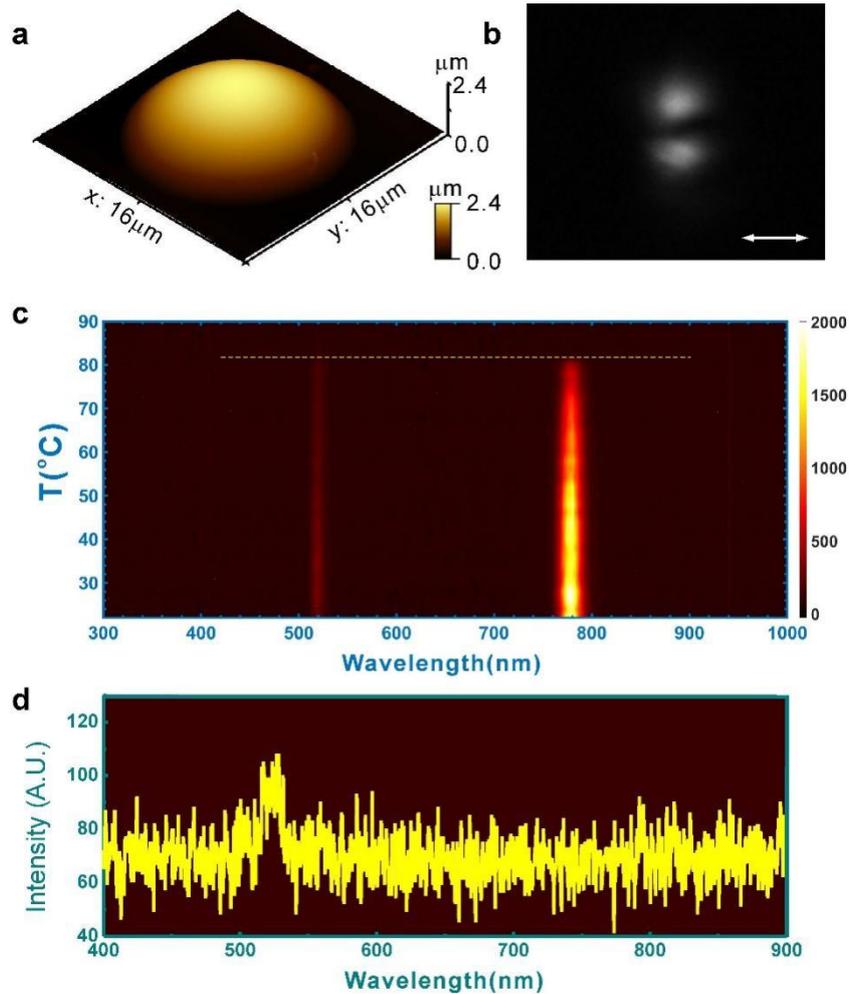

**Fig. 3 Nonlinear wavelength conversion in spherulites: a**, The AFM image of the spherulite. **b**, The transverse intensity distribution of the second harmonic generated in the spherulite illuminated with a linearly polarized fundamental wave. **c**, Nonlinear wavelength conversion as a function of temperature ranging between 22-90 °C. The yellow dashed line indicates that there is only THG above 82.6 °C. **d,** The nonlinear wavelength conversion spectrum at 84 °C, which shows only the THG.

Figure 3c also reveals the cascaded SFG process due to the mixing of the fundamental and second harmonic, resulting in the green light output that corresponds to the THG. Indeed, by heating the spherulite sample until the temperature above 82.6 °C, which has been determined to be the ferroelectric nematic-nematic phase change point by DSC, the output intensities at 516.6 nm and 775 nm are decreasing together. Above 82.6 °C (shown by the dashed line), the SHG disappears completely while the THG remains although at a very low-intensity level, as shown in Fig. 3d. This behavior clearly shows that the appearance of the THG below the phase-changing point is linked with SHG and thus the THG is mainly due to the SFG as well as the third order nonlinearity. Above the phase-changing point, there is no SHG signal since the material has lost the spontaneous polarity. We can still see a very weak THG output and this is from the third-order nonlinearity that is a universal effect in all materials. Also, we can conclude that the SFG output is much stronger than that of the THG due to the third-order nonlinearity since the former one is a second-order nonlinear effect.

By measuring the polarization change due to the anisotropy, as shown in supplementary Fig. S6, the difference between the anisotropic refractive indices ($n_\theta$-$n_r$) is determined to be 0.135. Moreover, we also notice that the spherulite is almost lossless in the visible and near-infrared wavelengths. The details can be found in Supplementary Fig. S7. With this strong anisotropy, the spherulite can work as a cylindrical half wave plate if the spherulite has the right thickness to create a $\pi$ phase difference between the radial ($E_r$) and the azimuthal ($E_\theta$) components of the E field.

Next, we note that the discovered structural vortex polarity resulting in a circular anisotropy in both linear and nonlinear optical properties of the spherulite matches perfectly the circular symmetry of the structured light beam with an orbital angular momentum (OAM) and/or cylindrical polarization. Therefore, next, we put forward a hypothesis that the spherulite can be used as a microscopic device to produce a series of optical vortices with multiple wavelengths, polarization states, and topological charges. Such structured light beams with multiple degrees of freedom in both space and time are likely to revolutionize the fields of classical and quantum communications and computing, enabling ultrafast, large-capacity, high-security state-of-the-art photonic systems[24-28]. To date, micro- and nano-scale sources of structured light that can be readily integrated on a chip have been mostly based on nanostructured engineered metamaterials or metasurfaces made using primarily top-down lithographic techniques. These techniques are usually time-consuming, and costly, and are subject to nanofabrication imperfections and defects that limit their real-world applications. The quest for bottom-up, e.g., self-assembly-based, nanofabrication techniques that are cost-effective, large-scale, and defect-free has been has been underway for decades. The organic spherulite crystal, synthesized in this work, offers a viable solution to the long-standing quest[29-34] enabling a compact, versatile approach for the generation of light with spin and orbital angular momenta, complex polarization states, and 'built-in' nonlinear frequency conversion capability.

Let's consider the incident left-hand circularly polarized electric field ($E_L$) that is a linear combination of the radially and azimuthally polarized vortex beams with topological charge 1 that can be written as

$$\boldsymbol{E}_L = (\boldsymbol{E}_r + i\boldsymbol{E}_\theta)\exp(i\theta). \qquad (1)$$

After it passes through the certain spherulite working as the cylindrical half waveplate, the $i\boldsymbol{E}_\theta$ component gaining an additional $\pi$ phase difference is flipped to $-i\boldsymbol{E}_\theta$ and thus the beam is transformed into a circularly polarized vortex with the opposite spin ($E_R$),

and the topological charge 2 as it follows from

$$\boldsymbol{E} = (\boldsymbol{E}_r - i\boldsymbol{E}_\theta)\exp(i\theta) = (\boldsymbol{E}_r - i\boldsymbol{E}_\theta)\exp(-i\theta)\exp(i2\theta) = \boldsymbol{E}_R\exp(i2\theta) \qquad (3)$$

Moreover, the above discussion can be extended to the domain of nonlinear optics of spherulites. For instance, due to the momentum conservation requirement, the SHG corresponding to the azimuthal component in the circularly polarized beam at the fundamental frequency with the topological charge 1 is also an azimuthally polarized beam with the topological charge 2. Furthermore, due to the interaction of such azimuthally polarized SHG with the azimuthal component of the fundamental wave, an azimuthally polarized OAM beam of topological charge 3 at the tripled frequency is expected as the result of the SFG. Figure 4a summarizes the results of the predicted OAM beam generation at the fundamental, doubled, and tripled frequencies in the spherulite structure.

To test the multiple vortex generations performance through a single spherulite, a circularly polarized beam at 1550nm from a femtosecond laser with its pulse width of 35 fs is incident onto a spherulite with a diameter of 30 μm and height of 5.2 μm. A Mach-Zehnder interferometer with a delay line is used to test the output beam at three different frequencies. The output beam at the fundamental frequency was extracted by a short-pass dichroic beamsplitter (SDBS) with its cutoff wavelength at $\lambda_{cutoff}$=1200nm and then interfered with the reference beam of right-hand circular polarization. The results are shown in Fig. 4b, which confirms the prediction of the generation of the right-hand circular polarized OAM beam with the topological charge of 2. With a quarter waveplate and a linear polarizer, we can also estimate the amount of the residual left-hand circular polarization and extract the polarization transformation efficiency of the spherulite to be 92% as shown in Supplementary Fig. S9.

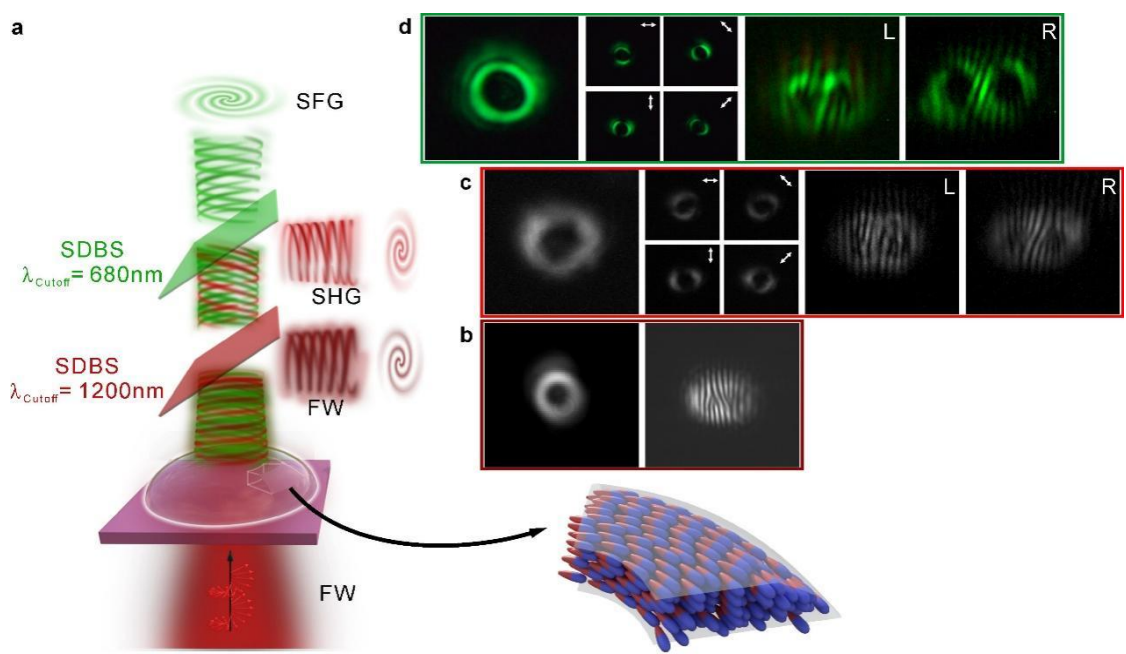

**Fig. 4 Characterization of the generated OAM beams at frequencies $\omega$, $2\omega$, and $3\omega$ from the spherulite. a** Schematic of the optical vortex (OAM beam) generation. The molecular dipoles are all well-aligned azimuthally. The left-hand circularly polarized wave at the fundamental frequency $\omega$ is incident onto the spherulite. The fundamental wave is then transformed into a right circularly polarized vortex of charge 2 upon interaction with the spherulite. The SHG shows azimuthal polarization of charge 2. The SFG shows azimuthal polarization of charge 3. **b**, The intensity and polarization measurements, and the off-center self-interference pattern for the output fundamental wave. **c**, The same measurements as in b for the output SHG beam: intensity profile after the rotating linear polarizer, and the off-center self-interference patterns with left and right circular polarizations separately. **d**, The intensity profile of

the SFG beam after the rotating linear polarizer and the off-center self-interference patterns with left and right circular polarizations separately.

We use the other SDBS with $\lambda_{cutoff}$ =680nm, which out-couples the SHG at 775nm. The polarization measurement performed on the SHG beam confirm the prediction that the second harmonic should be azimuthally polarized, as shown in Fig. 4c. The detection of the topological charge *l* OAM carried by an azimuthally polarized OAM beam can be demonstrated by exploring its two circularly polarized components with charges of *l*±1 as shown by Eq. (3)

$$\begin{pmatrix} -\sin\theta \\ \cos\theta \end{pmatrix} \exp(il\theta) = \frac{1}{2i}\left\{ \begin{pmatrix} 1 \\ i \end{pmatrix} \exp[i(l-1)\theta] - \begin{pmatrix} 1 \\ -i \end{pmatrix} \exp[i(l+1)\theta] \right\}. \qquad (3)$$

Here we used a quarter waveplate and a linear polarizer to extract the left- and right-circularly polarized components from the interference pattern, which shows charges 1 and 3, respectively. This implies that the second harmonics' charge is 2 and the polarization state is azimuthal. Finally, the third output beam transmitted through the second SDBS corresponds to the SFG at 516.6nm, as shown in Fig. 4d. By using the rotating linear polarizer technique at the output again, the polarization state of the SFG is also found to be azimuthal. An interferometric measurement reveals a fork with 2 tines with left circular polarization and with 4 tines with right circular polarization, which demonstrates that the SFG results in an azimuthally polarized optical vortex of charge 3.

**Conclusion**

The spontaneous vortex polarity of the synthesized spherulite crystal has been shown to give rise to circular anisotropic linear and nonlinear optical properties that in turn enable a unique platform for the optical vortex beam generations. The bottom-up, self-assembly-based, chemical approach was used to design an organic molecule with a large dipole moment that forms a circularly symmetric structure with vortex spontaneous polarity— a spherulite. Due to its structural symmetry, in the optical domain, the spherulite possesses circularly anisotropic refractive index and second-order nonlinearity, boosting various light-matter interactions, including spin-orbital coupling, SHG, and SFG, which facilitate a fundamentally new platform for structured light generation. Experimental studies supporting these predicted new regimes of light-matter interactions can be summarized as follows. When a left circularly polarized OAM beam with the topological charge 1 is transmitted through the spherulite, the beam is transformed into the right circularly polarized vortex of topological charge 2. Simultaneously, azimuthally polarized optical vortex beams of topological charge 2 at the doubled frequency and charge 3 at tripled frequency have been observed as a result of the SHG and SFG processes in spherulites, respectively. Thus, a single spherulite crystal enables a stand-alone micro-scale device that relies on the unique micro-scale spontaneous vortex polarity that has not been observed until now. Such a device is likely to enable future applications for high-dimensional quantum information processing, spatiotemporal optical vortices, and multi-dimensional optical communications.

**Methods**

**Morphology characterizations**

An Atomic Force Microscope (AFM) was used to characterize the absolute surface contour of the spherulites with different sizes. Using the piezo response force microscope (PFM) mode, we characterized the polarization response characteristics of the as-prepared spherulite. The AFM and PFM tests were carried out with the same equipment (Dimension FastScan, Bruker). In the PFM test, samples were on the ITO glass substrate.

We also characterized the surface morphology of the spherulite using a scanning electron microscope (GEMINISEM 500, ZEISS). Before observation, the sample was coated by sputtering a thin Pt film. The working voltage is 1kV to reduce the electron beam irradiation damage.

**Transmittance spectrum acquisition**

The transmittance spectra were acquired using the setup reported in our previous publication[35] equipped with a quartz tungsten-halogen lamp (SLS302, Thorlabs) and two spectrometers (HR2000+ES and NIRQUEST, Ocean Optics). The spherulite was designed to have an even thickness by sandwiching the sample between two glass slabs in the melting-cooling process. The anisotropic transmittance spectra of spherulite were obtained by using two linearly polarized beams as the incidence, i.e., azimuthal polarization and radial polarization.

**Data availability**

All the data supporting the claims of this paper are available from the corresponding authors upon reasonable request.

## Acknowledgments

This work was supported by the National Key R&D Program of China (Grant No. 2022YFB3806000). The Basic Science Center Project of NSFC (Grant No. 52388201). National Natural Science Foundation of China (Grant No. 11974203).

## Author contributions

Y.L. and J.S. developed the idea. L.Z. prepared all the chemicals. Y.L. did the spherulite fabrication. M.G. did the PFM characterizations. L.Y. and J.S. did the optical characterizations. J.S. and N.M.L wrote the paper. N.M.L., Y.W., Z.X., and J.M. contributed to the discussion. Y.S., J.Z., and J.S. supervised this work.

## Competing interests

The authors declare no competing interests.

## Supplementary information

### Supplementary Text 1. Preparation of the spherulite

The synthesized material was firstly dissolved by dichloromethane solvent (2 wt%) and then spin-coated into a film on a substrate. Then, the sample was heated to 130°C and held for 5min on a heating stage (MTI-250). During this process, the sample was fully melted and the dichloromethane solvent was volatilized totally. Finally, the sample was naturally cooled down to room temperature on the heating stage and the crystallized spherulite was formed on the substrate. During the cooling process, we can observe the change of the "Maltese cross", which also indicates the evolution of the crystal structure and the polarity order of the spherulite, as shown in Fig. S1.

We used glass and ITO glass as the substrates for the optical characterizations and the piezo response force microscope test. The substrate did not do any change on the spherulites crystallizations and the polarity order.

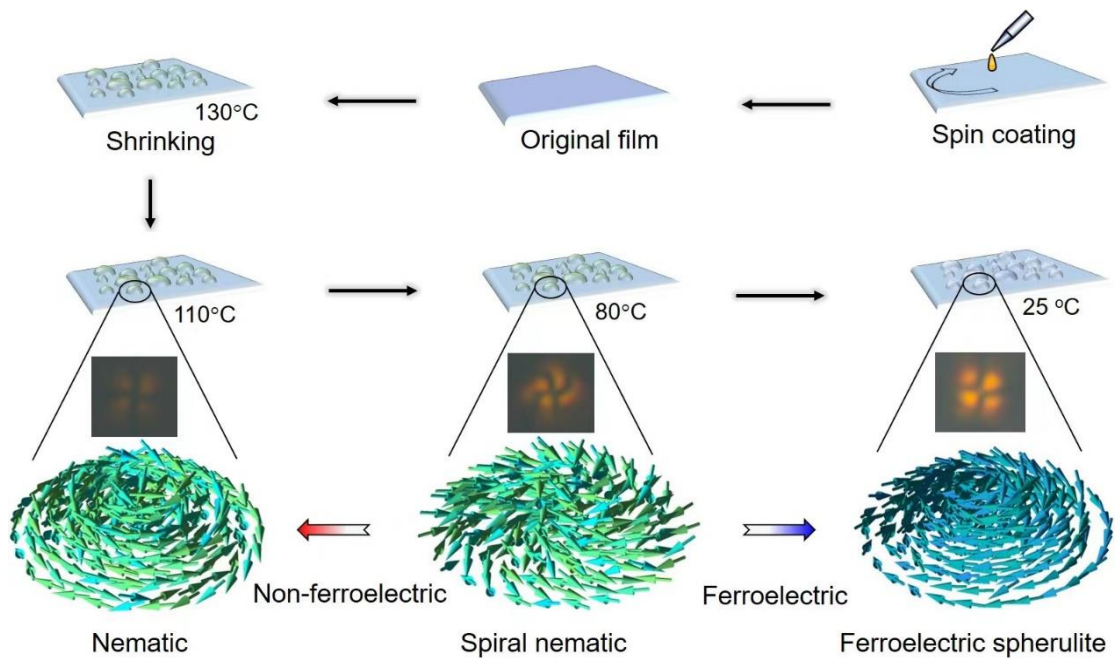

**Supplementary Fig. S1**. Preparation process of the spherulite.

## Supplementary Text 2. Surface contour of the spherulites

The white light interferometry (ZYGO NexView) test based on a nondestructive method with high efficiency was used to acquire the surface curvature of the spherulite. The spherulites were directly tested without any treatment and the measurement results can be found in Fig. S2, showing the geometric shape of each spherulite is spherical. By using Atomic Force Microscope (Dimension FastScan, Bruker), we can also determine the surface quality, the overall shape of the spherulites, as shown in Fig. S3, which tells us the surface roughness is rather small. Moreover, the surface contour measurements taken on 9 spherulites of different sizes (6 from white light interferometry and 3 from AFM) show that all these spherulites have the same shape, determined by the contact angle, which is 35°. This also demonstrates that our fabrication process is very reliable.

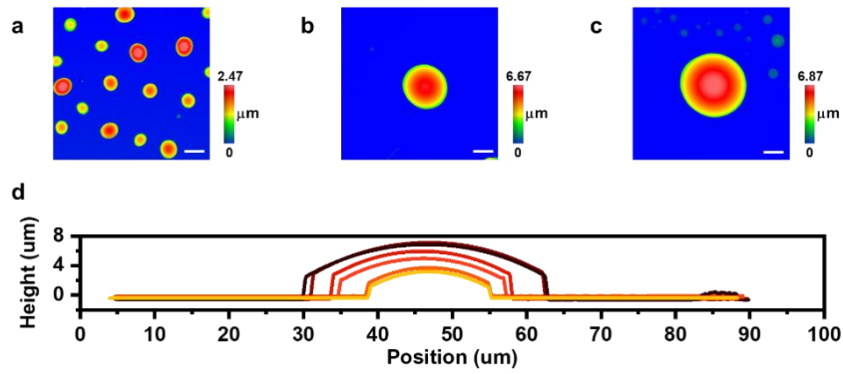

**Supplementary Fig. S2**. Surface contour of spherulites based on white light interferometry method. **a-c**, The thicknesses of spherulites with different feature sizes were characterized using a white light interferometry equipment. The scale bar in each panel is 10 μm. **d**, The curvature of the spherulites with different diameters based on the white light interferometry test.

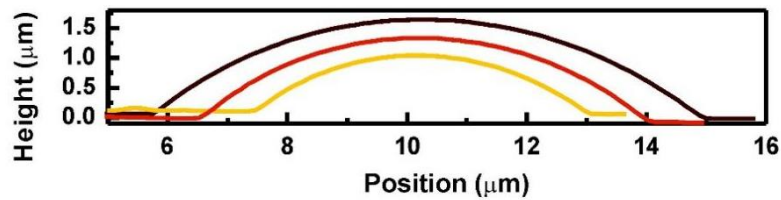

**Supplementary Fig. S3**. Surface profiles of the spherulites with different diameters based on the AFM characterization.

### Supplementary Text 3. Differential scanning calorimetry (DSC) test

Thermal property of the sample was characterized utilizing a DSC equipment (TA DSC-Q2000). The phase transition temperatures were obtained in quasi-equilibrium conditions under cooling process at a rate of 5°C/min by DSC. The DSC curve in Fig. S5 reveals the ferroelectric nematic to nematic phase transition at 82.6°C and 126.9°C should be the clearing point. When the temperature is higher than 82.6°C, the material loses the spontaneous polarity and thus no SHG anymore, which agrees well with the case in Fig. 3c.

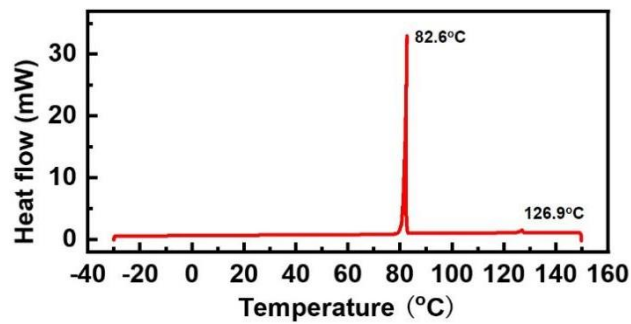

**Supplementary Fig. S4**. DSC profile of a cooling process at the rate of 5°C/min. The initial mass of the sample is 7.48 mg.

## Supplementary Text 4. Calculation of the second order nonlinear coefficient

The second order nonlinear polarization coefficient $d_{eff}$ can be determined by fitting the intensities of the fundamental wave $I(\omega)$ and the SHG $I(2\omega)$ by using the equation below:

$$I(2\omega) = C \frac{1}{n^2(2\omega)} d_{eff}^2 l^2 I^2(\omega) \tag{S1}$$

where $l$ is the effective thickness. $n(2\omega)$ is the refractive index at 775 nm and is equal to 1.8. C is an equipment related constant and firstly calibrated by a standard $LiNbO_3$ single crystal film. We also do the fitting regarding to the relation between the input power and the output power.

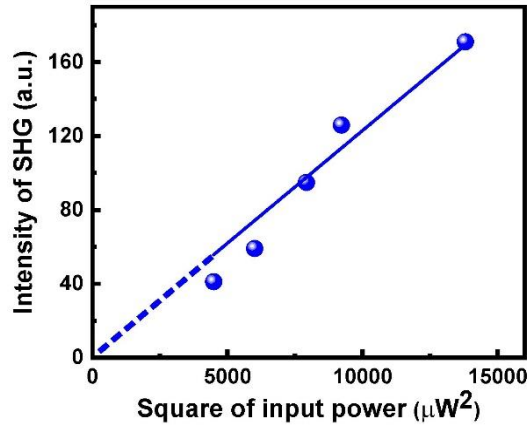

**Fig. S5**. Fitting on the relation between $I(2\omega)$ and $I(\omega)^2$.

## Supplementary Text 5. Δn determination of the spherulite

A more precise way to determine the level of the anisotropy Δn was used by measuring the polarization change when a linearly polarized beam with its E-field polarized at an angle of $\theta = 45°$ with respect to horizontal axis passed through a flat spherulite, which was prepared on purpose. In the experiment, the thicknesses of the flat spherulite were measured by a step profiler (Bruker DektakXT). The test region is far from the center of the spherulite so that the local anisotropy is also able to be expressed as $n_x$ and $n_y$ in a Cartesian coordinate. We prepared two flat spherulite samples with different thicknesses: 5.2 μm and 6.3 μm. A CW laser at 1550 nm was used as the incident beam. After the sample, we used a linear polarizer and a power meter to detect the intensities along or perpendicular to the polarizer oriented at an angle of α ranging from 0-180° with a step of 10°, as shown in Fig. S6. The intensities of each orientation are varying

with α according to the equations below:

$$I_{\parallel} = A(-\sin\alpha + \exp(i\frac{2\pi}{\lambda}\Delta nd)\cos\alpha)^2 \quad (S2a)$$

$$I_{\perp} = A(\cos\alpha + \exp(i\frac{2\pi}{\lambda}\Delta nd)\sin\alpha)^2 \quad (S2b)$$

where A is intensity of the incident beam along x or y direction. By fitting the theoretical calculation to the measurement results in Fig. S6 b and c based on Eq. S2, the Δn was determined to be 0.135.

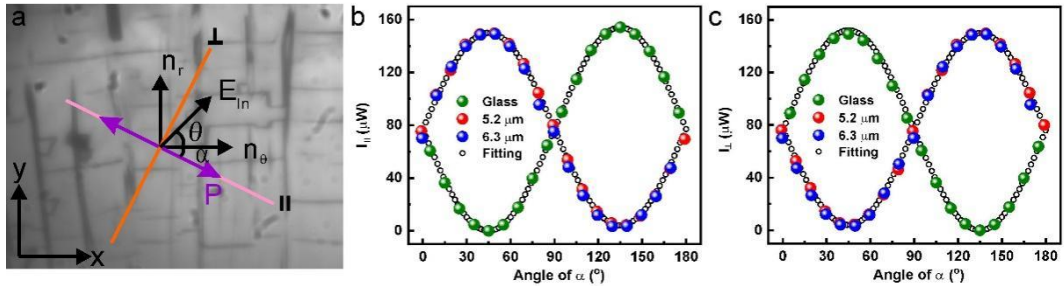

**Supplementary Fig. S6**. Fitting results of the polarization change with samples of two different thicknesses. **a**, Infrared optical image of the flat sample. **b**, Intensities along the polarizer. **c**, Intensities perpendicular to the polarizer.

### Supplementary Text 6. Transmittance spectra

The transmittance spectra were also measured on a flat spherulite. The spherulite shows high transmittance from 500 nm-1600 nm at both azimuthal and radial polarizations with negligible extinction loss.

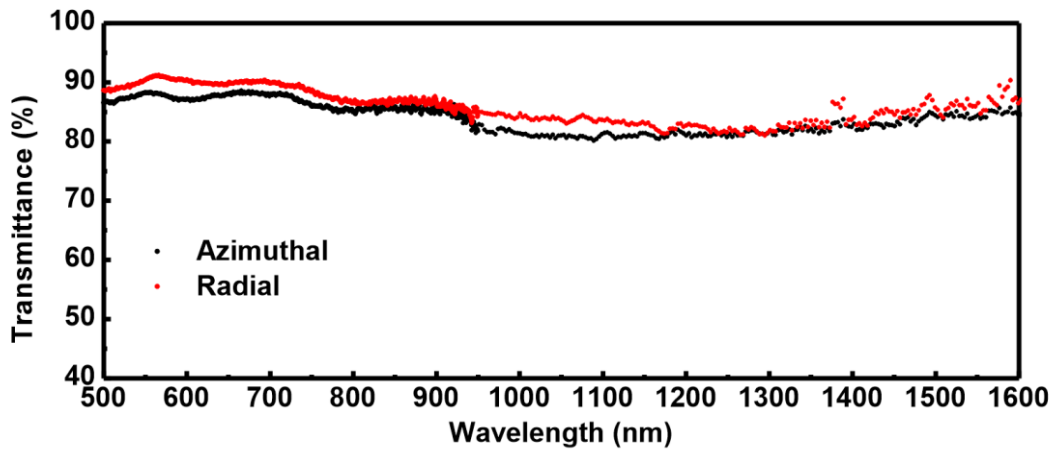

**Supplementary Fig. S7**. **Anisotropic optical property of the spherulite.** Transmittance spectra of the incident polarization along the azimuthal and radial direction of the spherulite film.

### Supplementary Text 7. Theoretical prediction of the curvature effect on the polarization conversion efficiency of the fundamental wave

The spherulite sample we used to generate the optical vortex beams was in spherical-cap shape whose contact angle was 35º. In principle, we need a certain thickness so that the anisotropic index can bring in a π phase delay between the radial and azimuthal polarizations. However, the spherical surface may result in a weak focusing effect as well as a deviation from the π phase difference, both of which may determine conversion efficiency from the left circular polarization to the right circular polarization regarding to the fundamental-wave. Obviously, this effect is related to the beam size of the incidence and the overall size of the spherulite. A small beam size can only cover the center of a large size spherical-cap, where the surface is approximately flat and thus a much better conversion efficiency. Here we performed a theoretical calculation to predict the conversion efficiency regarding to the beam size and the radius of the spherulite and the result is plotted in Fig. S8. In the experiment, the incident beam is focused to be 8 μm onto a spherulite with a radius of 30 μm and the theoretical conversion efficiency is 97%.

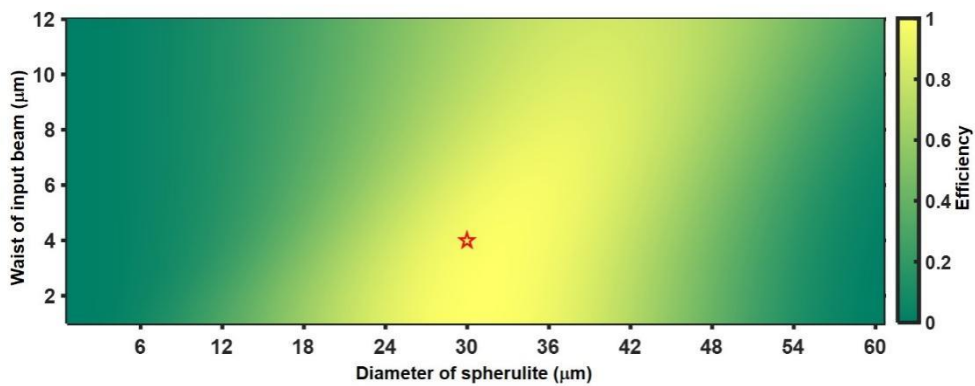

**Supplementary Fig. S8**. Theoretical analysis of the fundamental wave vortex generation efficiency. The theoretical efficiency in our case is labeled by the star.

## Supplementary Text 8. Conversion efficiency of the vortex generated by fundamental wave

The optical images of the output beam right after the spherulite are shown in Fig. S9. By calculating the overall intensities of the output beams with left circular polarization and right circular polarization., the efficiency of the optical vortex generation with charge 2 is determined to be 92%.

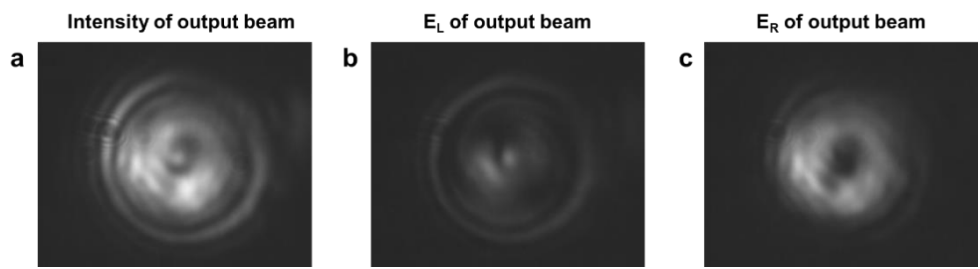

**Supplementary Fig. S9**. Intensity distribution of the output beam. (a) Total intensity. (b) Intensity of the $E_L$. (c) Intensity of the $E_R$.